\documentclass[a4paper,11pt]{scrartcl}
\usepackage{ILD}
\usepackage{tagging}
\usetag{ILD}
\usepackage{subcaption}
\usepackage{hyperref}
\usepackage{wasysym}
\usepackage{amssymb}
\usepackage[backend=biber,style=numeric-comp,sorting=none,mcite=true,doi=false,subentry]{biblatex}
\usepackage{slashed}
\usepackage{wrapfig,rotating}
\usepackage{pgfpages}
\usepackage{fancybox,graphicx}
\usepackage{graphbox}
\usepackage{epstopdf}
\AppendGraphicsExtensions{.eps.gz}
\title{
  $\stau$ searches at future e$^+$e$^-$ colliders

}
\ildproc{phys}{2024}{007}
\localrep{DESY-24-157\\}
\date{\today}

\addauthor{
  \untagged{webofc}{
Mikael Berggren\institute{1},
}
\tagged{webofc}{
  \firstname{Mikael} \lastname{Berggren}\inst{1}\fnsep\thanks{Speaker,\email{mikael.berggren@desy.de}}, \firstname{Maria Teresa} \lastname{N{\'u}{\~n}ez Pardo de Vera}\inst{1}, \firstname{Jenny} \lastname{List}\inst{1}
  
  \tagged{POS,webofc}{On behalf of }
the ILD concept group.
\tagged{webofc}{\thanks{
Contributed to The International Workshop on Future Linear Colliders (LCWS2024) 
 8-11 July 2024,
Tokyo, Japan

}
}

  }

  \untagged{webofc}{
Maria Teresa N{\'u}{\~n}ez Pardo de Vera\institute{1},
}
\tagged{webofc}{
  \firstname{Mikael} \lastname{Berggren}\inst{1}\fnsep\thanks{Speaker,\email{mikael.berggren@desy.de}},\firstname{Maria Teresa} \lastname{N{\'u}{\~n}ez Pardo de Vera}\inst{1},\firstname{Jenny} \lastname{List}\inst{1}

  }

  \untagged{webofc}{
Jenny List\institute{1}
}
\tagged{webofc}{
  \firstname{Mikael} \lastname{Berggren}\inst{1}\fnsep\thanks{Speaker,\email{mikael.berggren@desy.de}},\firstname{Maria Teresa} \lastname{N{\'u}{\~n}ez Prado de Vera}\inst{1},\firstname{Jenny} \lastname{List}\inst{1}

  }

}

\addinstitute{1}{
    Deutsches Elektronen-Synchrotron DESY,\\
  Notkestr. 85, 22607 Hamburg, Germany

}
\onbehalfof{
    
}

\abstract{
  The direct pair-production of the superpartner of the $\tau$-lepton, the $\widetilde{\tau}$,
is one
of the most interesting channels to search for SUSY in:
the $\widetilde{\tau}$ is
likely to be the lightest of the scalar leptons,
and is one of the  most experimentally chalanging ones.
The current model-independent $\widetilde{\tau}$ limits come from LEP,
while limits obtained at the LHC do extend to higher masses, but are model-dependent.
The future Higgs factories will be powerful facilities for SUSY searches, offering advantages
with respect to previous electron-positron colliders as well as to hadron machines.
In order to quantify the capabilities of these future $e^+e^-$ colliders,
the ``worst-case'' scenario for $\widetilde{\tau}$ exclusion/discovery has been studied, taking
into account the effect of the $\widetilde{\tau}$ mixing on $\widetilde{\tau}$ production
cross-section and detection efficiency.
To evaluate the latter, the ILD concept, originally developed for the International
Linear Collider (ILC), and the ILC beam conditions at a centre-of-mass energy of
$500$\,GeV have been used for detailed simulations.
The obtained exclusion and discovery reaches extend to only a few GeV below the kinematic limit
even in the worst-case scenario.

The results of the detailed simulation study are then discussed in view of the experimental environment
of other proposed Higgs factory projects.

}

\titlecomment{%
Contributed to The International Workshop on Future Linear Colliders (LCWS2024) 
 8-11 July 2024,
Tokyo, Japan

}

\def\leqsim{\mathbin{\;\raise1pt\hbox{$<$}\kern-8pt\lower3pt\hbox{$\sim$}\;}}
\def\geqsim{\mathbin{\;\raise1pt\hbox{$>$}\kern-8pt\lower3pt\hbox{$\sim$}\;}}

\def\p#1{\mbox{$ \mbox{\bf p}_1                                         $}}

\newcommand{\stau}    {\mbox{$ \tilde{\tau}                                $}}

\newcommand{\eeto}    {\mbox{$ {\, \mathrm e}^+ {\mathrm e}^- \to             $}}

\newcommand{\dgree}   {\mbox{$ ^\circ                                      $}}

\newcommand{\ba}{\begin{array}}
\newcommand{\ea}{\end{array}}
\newcommand{\bc}{\begin{center}}
\newcommand{\ec}{\end{center}}
\newcommand{\be}{\begin{eqnarray}}
\newcommand{\eeq}{\end{eqnarray}}
\newcommand{\bes}{\begin{eqnarray*}}
\newcommand{\ees}{\end{eqnarray*}}
\newcommand{\Kz}{\ifmmode {\rm K^0_s} \else ${\rm K^0_s} $ \fi}
\newcommand{\Zz}{\ifmmode {\rm Z^0} \else ${\rm Z^0 } $ \fi}
\newcommand{\xxbar}{\ifmmode {\rm x\bar{x}} \else ${\rm x\bar{x}} $ \fi}
\newcommand{\rphi}{\ifmmode {\rm R\phi} \else ${\rm R\phi} $ \fi}

\def    \missEt      {\ifmmode{/\mkern-11mu E_t}\else{${/\mkern-11mu E_t}$}\fi}
\def    \missE       {\ifmmode{/\mkern-11mu E}\else{${/\mkern-11mu E}$}\fi}
\def    \missp       {\ifmmode{/\mkern-11mu p}\else{${/\mkern-11mu p}$}\fi}
\def    \misspt      {\ifmmode{/\mkern-11mu p_t}\else{${/\mkern-11mu p_t}$}\fi}

\begin{document}
\titlepage

\section{Introduction}

  The standard model (SM) works excellently - but there are problems.
  On one hand, there are theory-experiment discrepancies such as
  the value of the magnetic moment on the muon ($g-2$) which shows a close to 5~$\sigma$ 
  discrepancy \cite{Muong-2:2023cdq,Aoyama:2020ynm}.
  There are anomalies in the flavour sector \cite{HeavyFlavorAveragingGroup:2022wzx}, 
  and possibly
  on the value of $M_W$ \cite{CDF:2022hxs}.
  Also, the SM lacks explanations for observed phenomena:
  Dark matter certainly exists, and the current-day acceleration of the expansion of the universe indicates
  the existence of dark energy.
  There is the issue of naturalness and the hierarchy problems: Why is the Higgs mass so small, 
  and why does it remains so, when nothing in the SM seems to forbid very large quantum 
correction from loops?
  The coupling constants of the fundamental interactions seem to tend to a single value,
  but they do not actually unify at the same scale.
  There is no reason in the SM that electric charge should be quantised, but it 
  clearly is.
  In the SM, the cosmological constant is wrong by 120 orders
      of magnitude.

  All these issues point to the need for some physics beyond the SM (BSM).
Among the few internally consistent models for BSM, super-symmetry
(SUSY)
\cite{Martin:1997ns,Wess:1974tw,Nilles:1983ge,Haber:1984rc,Barbieri:1982eh}, 
stands out as a prime candidate that offers solutions and/or hints to solutions to
several of the problems, including
the naturalness and the hierarchy problems,
 the coupling constant unification at an unique GUT scale,
and an explanation for the quantisation of charge.
It can also provide a candidate for  Dark Matter,
and an explanation of the observed value of $g-2$ of the muon.
The fact that the cosmological constant is very small, but not vanishing, can also be understood
in some versions of SUSY.

No clear signal of SUSY has been seen in the data from the LHC so far, nor
did searches at LEP-II find any indications of SUSY. 
This has lead to a sentiment in the community that SUSY is strongly
challenged.
In fact, what is strongly
challenged is the cMSSM (aka mSUGRA) paradigm that was popular pre-LHC.
This paradigm contains a minimal number of parameters,
and couples the electroweak and strong sectors of SUSY closely,
and thus predicted that coloured states (the squarks and the gluino) should
be in reach of the LHC. 
These have now been excluded up to masses well above 1 TeV.
But this coloured sector has little bearing on the issues mentioned above -
the issues only require rather light and close-together electroweak states to exists.
In fact, the precision electroweak measurements at LEP predicted that
the Higgs mass should be less than 140 GeV if SUSY was assumed \cite{Djouadi:2005gj}, while
a much larger value of 285 GeV would have been allowed by the SM alone~\cite{ALEPH:2005ab},
and indeed, a Higgs was observed below the SUSY-imposed limit.
Both LEP and LHC have observed an excess of Higgs-like events
at around 95 GeV, which could be a sign of a second scalar Higgs,
required to exist in SUSY, but not in the SM.
Both ATLAS and CMS observes an persistent excess of events that can
be interpreted as Chargino/Neutralino production at a mass of around
200 GeV and a mass-difference to the LSP of around 20 GeV \cite{ATLAS:2021moa,ATLAS:2019lng,CMS:2021edw,CMS-PAS-SUS-23-003}.
While some specific models can be excluded by the LHC,
a full scan of the 18 parameters of R-parity and CP conserving SUSY
recently performed by ATLAS shows that hardly any points in the
parameter-plane beyond what was probed by LEP-II can be excluded \cite{ATLAS:2024qmx}.
The reason why LEP could conclusively exclude SUSY almost up to the
kinematic limit, while the LHC cannot, is that the blessing of the
high production cross-section for strong processes becomes a curse
if the signal is colour-neutral: no increase of the signal from
strong production, only of the background.

Therefore, a lepton collider with an energy well above the energy of LEP-II
will be paramount to be able to further exploit the SUSY parameter-space
in a model independent way. The proposed Higgs Factory can fill this role,
in particular if it is designed to reach energies up to the TeV range,
as the different proposals for linear colliders are. Among them,
 the International Linear Collider (ILC) \cite{ILCInternationalDevelopmentTeam:2022izu, Behnke:2013xla, ILC:2013jhg, Adolphsen:2013jya, Adolphsen:2013kya, Behnke:2013lya}
 was proposed as a mature option 
for the future $e^+ e^-$
Higgs factory, and is the main option used for this study.
The baseline running scenario assumes starting at a centre-of-mass energy of
250 GeV followed by a 500 GeV stage and 1 TeV considered as the possible upgrade. 
In the
assumed 22-year running period the ILC is expected to deliver the integrated luminosities of about
2 ab$^{-1}$ at 250 GeV and 4 ab$^{-1}$ at 500 GeV, with an additional 200 fb$^{-1}$ collected at the top-quark 
pair-production threshold around 350 GeV~\cite{Barklow:2015tja}. 
The design includes polarisation for both $e^-$ and $e^+$ beams, of
80\% and 30\%, respectively, which is the unique feature of the ILC.
Other Higgs factories are also touched upon: The Compact Linear Collider (CLIC)~\cite{Brunner:2022usy, CLICdp:2018cto, Linssen:2012hp}
 and the Cool Copper Collider (C$^3$)~\cite{Vernieri:2022fae},
both linear and capable to reach the TeV regime, 
and the Future Circular Collider, $e^+ e^-$ version (FCCee)~\cite{Bernardi:2022hny, FCC:2018byv, FCC:2018evy}
and the Circular Electron Positron Collider (CepC)~\cite{Gao:2022lew, CEPCPhysicsStudyGroup:2022uwl, CEPCStudyGroup:2018rmc, CEPCStudyGroup:2018ghi},
dedicated Higgs Factories, reaching at most the top-threshold.


  \begin{figure}[t]
  \centering
       \includegraphics [scale=0.38]{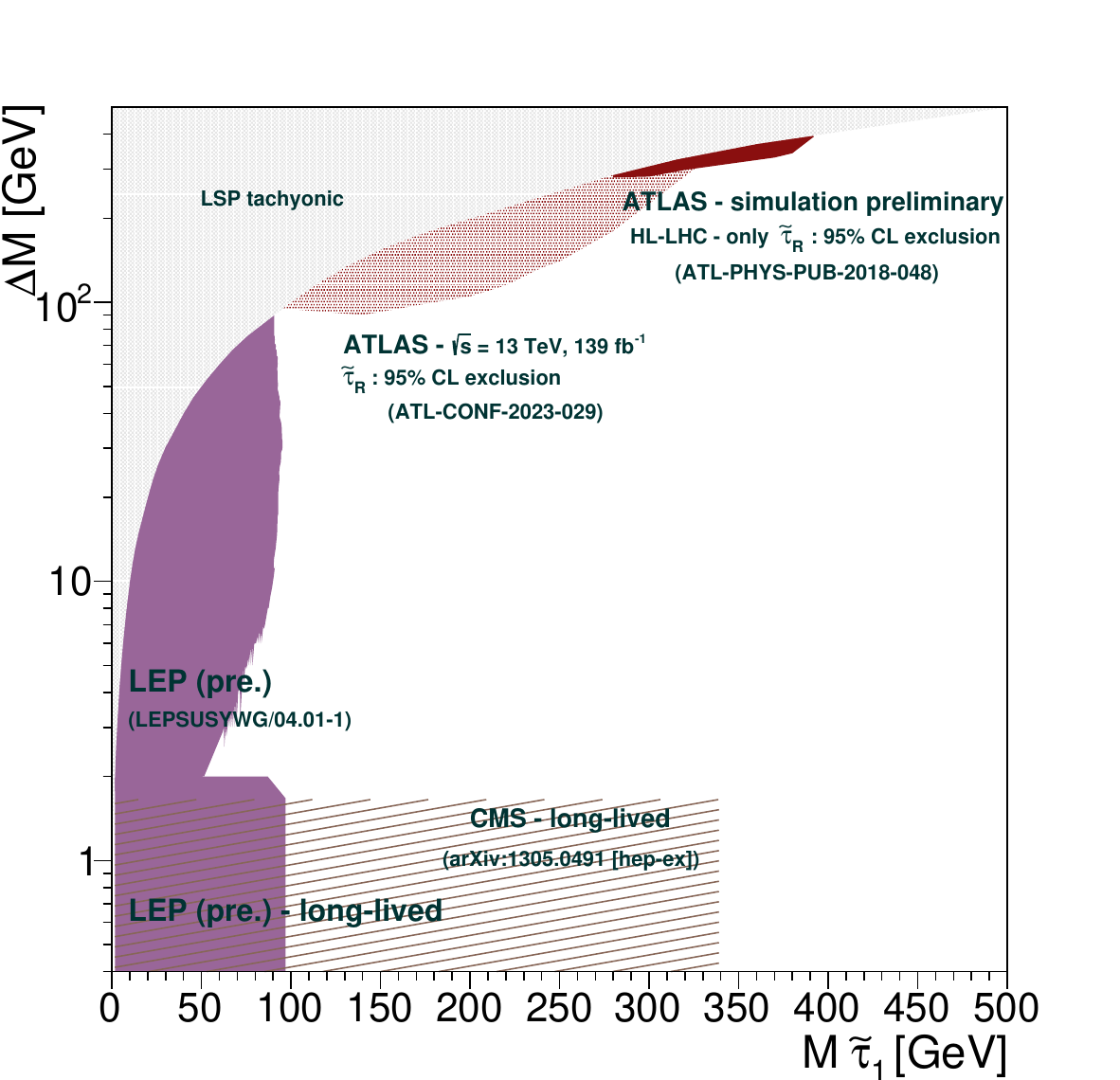}
  
       \caption{Limits in the $\Delta M$ vs. $M_{\stau}$ plane, from the LEP combination,
  from the ATLAS-experiment at the LHC, and the HL-LHC projection from ATLAS.
         \label{fig:currentlimits}}
\end{figure}

  \subsection*{\textit{ Motivation for $\stau$ searches. Current limits}}
 For SUSY searches it is a good idea
 to search for well motivated and maximally 
difficult Next-to-Lightest SUSY Particles (NLSPs):
If one can find this, then one can find any other NLSP.
  The $\stau$
  has two weak hypercharge eigenstates (${\stau}_R , {\stau}_L$),
  which are not mass degenerate.
  Mixing yields the physical states  (${\stau}_1 , {\stau}_2$),
  the lightest one being
    likely to be
    the lightest sfermion, due to the stronger trilinear couplings
    expected for the third family SUSY particles.
   If R-parity is assumed to be conserved,
    the $\stau$ will be pair-produced in the s-channel via
    $Z^0/\gamma$ exchange.
    The production cross-section can be quite low,
    since  $\stau$-mixing can suppresses the coupling
    to the $Z^0$ component of the neutral current, so that only $\gamma$
    exchange contributes.
    The $\stau$ will decay to the LSP and a $\tau$, implying a more
    difficult signal to identify  than that of other sfermions,
   since the $\tau$ decays partially invisibly. In addition,
  mixing can further reduce detectability.
 Furthermore, the presence of a $\stau$ close in mass to the LSP
  can contribute to co-annihilation between the two in the early universe,
  and in this way avoid an over-abundance of SUSY WIMP dark matter~\cite{Ellis:1998}.
  Finally, the $\stau$ is the SUSY particle least constrained from
  current data.
We see that the $\stau$ satisfies both conditions: it is both a well motivated and maximally 
difficult NLSP candidate.

    Figure \ref{fig:currentlimits} shows the current limits for the $\stau$ in the
    plane of $\Delta M$ vs. $M_{\stau}$, together with a projection
    of the expected results at the high luminosity phase of the LHC (HL-LHC).
    The LEP limit is valid for any mixing and any values of the unshown parameters.
    This is from the unpublished LEP combination~\cite{LEPSUSYWG/04-01.1}.
    The 
     PDG~\cite{Workman:2022ynf} quotes the best published limit (from 
     DELPHI~\cite{Abdallah:2003xe}) of  81.9 GeV
     for any mixing if $\Delta M > $ 15 GeV), and 26.3 GeV for any mixing and any $\Delta M$.
        The ATLAS limit~\cite{ATLAS:2023conf} is model dependent; it is for a pure $\stau_R$.
       It only  excludes  very high $\Delta M$, where it is
    unlikely that the $\stau$  would be the NLSP.
    No discovery potential is expected.
    The HL-LHC projection expects to be able to exclude somewhat higher $\stau_R$ masses,
    but still for very high $\Delta M$, and with  no discovery potential~\cite{ATLAS:2018diz}.

\begin{figure}[t]
  \centering
   \includegraphics [scale=0.22]{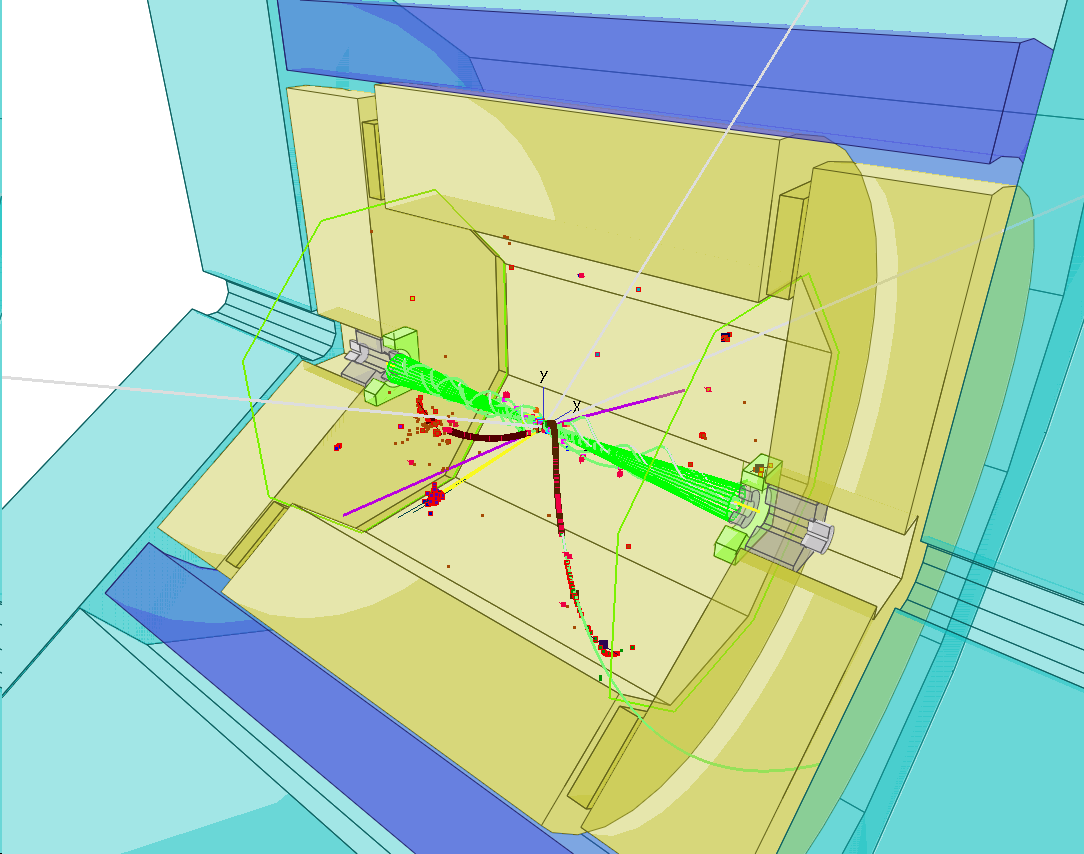}
    \caption{A $\widetilde{\tau}$ event at the ILC operating at $\sqrt{s}$ = 500 GeV, fully simulated in the ILD detector.
      $M_{\widetilde{\tau}}$ = 230 GeV, and $\Delta M =$~10~GeV.}
  \label{fig:anevent}
\end{figure}

\section{$\stau$ properties at $e^+e^-$ colliders}
Figure \ref{fig:anevent} shows a typical fully simulated $\stau$ event in the ILD detector.
This event well illustrates the expected properties of the signal.
There will be large missing energy and momentum, due both to the
undetected LSPs and to the neutrinos.
A
large fraction of detected activity in central detector,
since the $\stau$'s are scalar particles and hence  isotropically produced.
Once again due to the unobserved LSPs, there will be a
large angle between the two $\tau$-lepton directions,
also leading to unbalanced transverse momentum.
Contrary to many backgrounds, no forward-backward asymmetry is expected.
The SM background to a signal with these properties will be processes with
real or fake missing energy.
On one hand, there are irreducible backgrounds, namely 
four-fermion production with two of the fermions being neutrinos and two $\tau$'s.
On the other hand there will be ``almost'' irreducible processes, e.g.
$\eeto \tau\tau$, $ZZ \rightarrow \nu\nu ll$ , or $WW \rightarrow l \nu  l \nu$ ($l = e$ or $\mu$)
i.e. processes with real missing energy, and visible systems that can easily be mistaken
as decay-products of $\tau$'s,
and also events where part of the final state is outside the acceptance of the detector -
corresponding to the inevitable holes at very low angles
for the in- and out-going beam-pipes, e.g.
  $\eeto \tau\tau + ISR$,  t-channel $\eeto \tau\tau ee $, or $\gamma\gamma \rightarrow \tau\tau$.



\begin{figure}[b]
          \centering
       \includegraphics [scale=0.28]{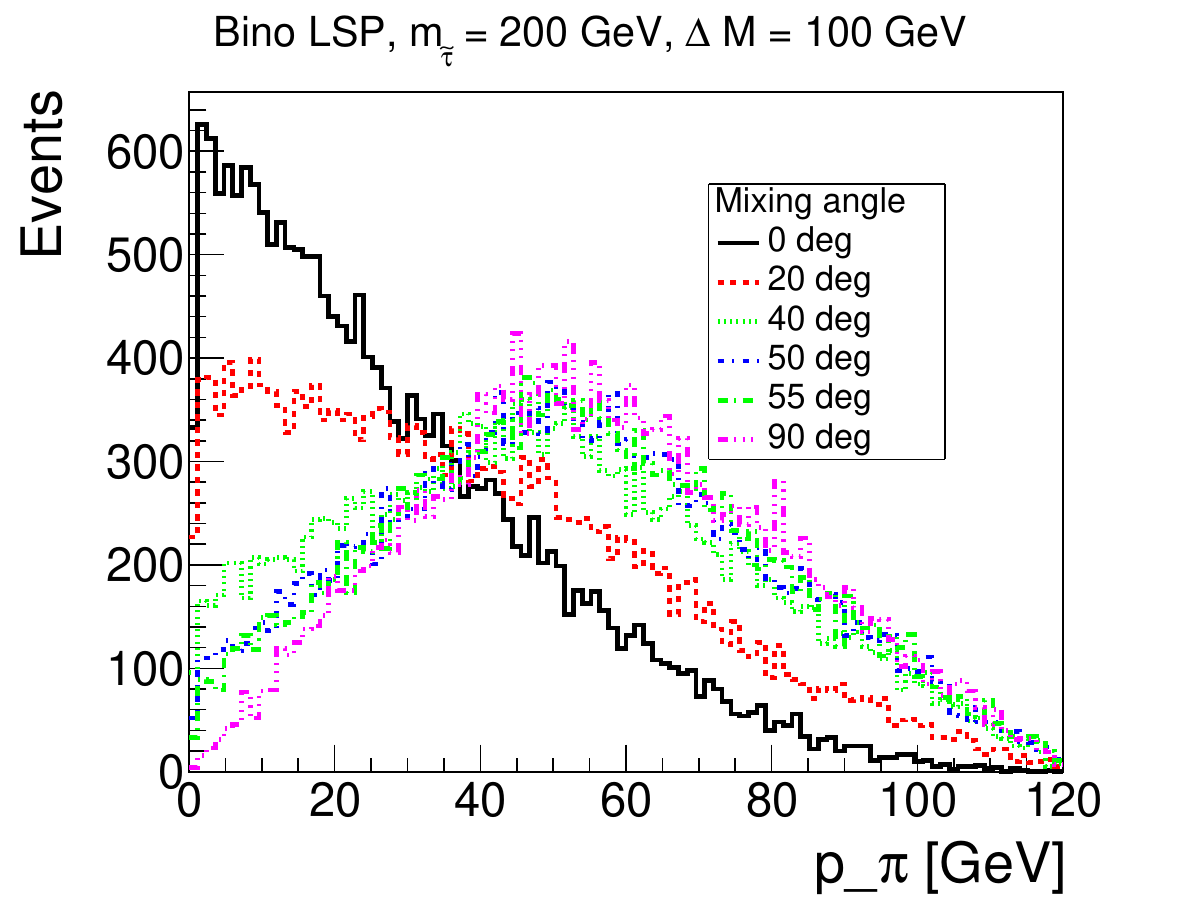}
      \includegraphics [scale=0.35]{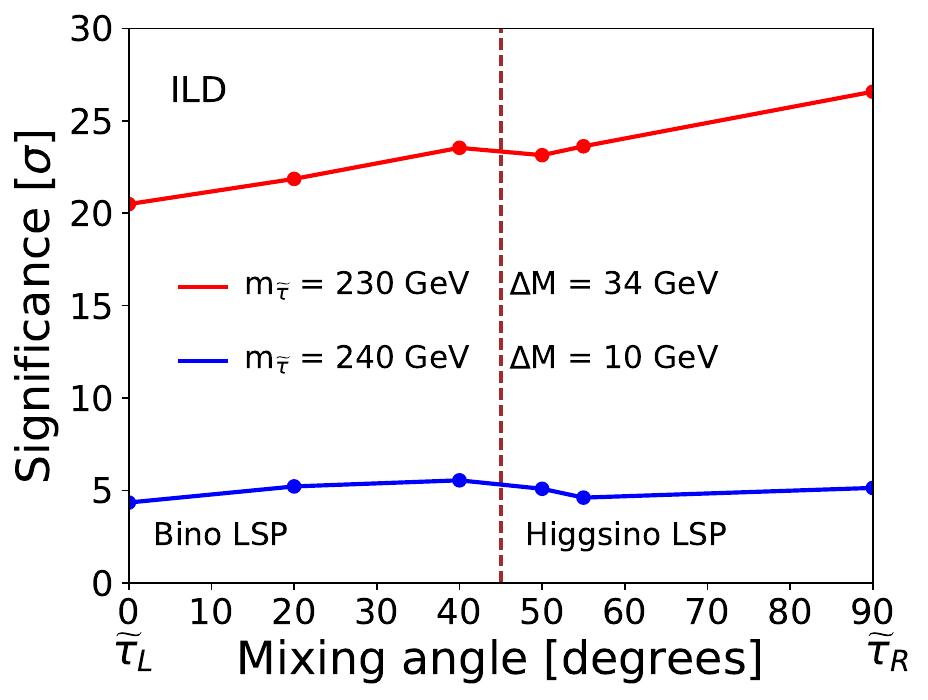}
          \caption{Left: Momentum distribution of the pions coming from ${\tau}$-decays for different $\widetilde{\tau}$ mixing angles.
      The neutralino was taken to be pure bino. 
Right: Signal significance weighting both polarisations using the likelihood ratio statistic in the H20 ILC conditions. }
          \label{fig:mixingsetc}
        \end{figure}

   As already pointed out,  the production cross-section depends on mixing.
   But, in addition, the  visibility of the signal also depends on mixing,
   since the $\tau$ polarisation influences the visible spectrum of the $\tau$
   decay products, see Figure \ref{fig:mixingsetc} (left),
       and  $\tau$ polarisation depends on both the $\stau$ and the
       neutralino mixing angles.
     Therefore, to make sure that one studies the worst case, the combination of
        low cross-section and low visibility should be found.
 At the ILC, both beams are polarised, and same
 luminosity will be collected for LR and RL beams.
 This allows to 
   use the Likelihood-ratio statistic to weight both polarisations, \textit{ viz.}
\begin{equation}
   N_{\sigma} = \frac{\sum^{n_{samp}}_{i=1} s_i \ln{(1+s_i/b_i)}}{
      \sqrt{ \sum^{n_{samp}}_{i=1} n_i \left [\ln{(1+s_i/b_i)} \right]^2}}
\end{equation}
 where $s_i$ and $b_i$ is the expected signal and background in sample $i$. $n_i$ is either $s_i +b_i$ (exclusion),
  or $b_i$ (discovery), and $n_{samp}$ is the number of distinct samples.
   Using this statistic results in a sensitivity that is almost uniform with respect to the 
   mixing angles, with a slight minimum at $\sim 55^\circ$,
   as can be seen in Figure  \ref{fig:mixingsetc} (right).
\section{ILD full simulation analysis}
The International Large Detector (ILD) concept~\cite{ILDConceptGroup:2020sfq}  
is used as the detector in this study.
The main tracker of ILD is a large TPC, offering excellent pattern recognition and 
particle identification capabilities 
as well as very good momentum resolution, with a minimal material budget.
Inside the TPC, closest to the interaction point, a silicon pixel vertex detector allows to reach impact-parameter
measurement precision of 5 $\mu$m, and outside the TPC, a large silicon strip detector helps to further enhance the
momentum resolution down to $\sigma(1/P_T)$ = 1 $\cdot 10^{-5}$. The highly granular electromagnetic and hadron calorimeters
are both placed inside the 3.5 T superconducting solenoid, and the return yoke is instrumented to detect muons.
The low angle region is of utmost importance for this analysis. Here, 
a set of discs of silicon detectors allows to reconstruct charged tracks down to 7$\dgree$ from the beam axis.
Below this angle, the forward calorimeters are placed: The luminosity monitor, LumiCal, behind it the low angle hadron
calorimeter, LHCal, which assures that also hadrons can be detected to the lowest angles. 
In the very forward region the BeamCal is placed,
mounted directly on the beam-pipe. The holes in the BeamCal for the beam-pipes are the only uninstrumented part of the system,
and represents an angle to the beam of 6 mrad. 
  This study uses the IDR 500 GeV FullSim samples~\cite{ILDConceptGroup:2020sfq,Berggren:2021sju},
 covering the full SM background with all
$e^+e^-$/$e^{+/-} \gamma$/$\gamma\gamma$ processes ($>10^7$ events).
The ILC  beam-spectrum and pair background were calculated and generated with 
GuineaPig~\cite{Schulte:1999tx}, and low $P_T$ hadrons
from a dedicated generator~\cite{Chen:1993dba}.

For the signal, the mass-spectrum was 
 obtained with Spheno~\cite{Porod:2011nf}, and the events were generated with Whizard~\cite{Kilian:2007gr}.
  The detailed fast simulation SGV~\cite{Berggren:2012ar} 
with the ILD geometry was used for detector simulation and high-level reconstruction.
  The pair background and low $P_T$ hadrons were extracted from FullSim, and added to the SGV-produced events.
  10000 events per point and polarisation were generated, at 
  1867 mass-points, resulting in a total of 37 $ \times 10^6$ events.
\subsection*{\textit{ Event selection}}
The event selection chain starts by selecting properties $\stau$-events at any given mass-point must have.
The missing energy must be at least twice the LSP mass,
and the
visible mass must fulfil  $M_{vis} < 2 \times  (M_{\stau} - M_{LSP})$ GeV.
Furthermore, there should be two well identified $\tau$'s and little other activity
and the higher of the two jet momenta should be below the highest value kinematically allowed at
the studied mass-point, \textit{ viz.}
\begin{equation}    
    P_{max}= \frac{E_{beam}}{2} \left [ 1 -
        \left ( \frac{M_{LSP}}{M_{\stau}} \right )^2 \right ]
          \left [1 + \sqrt{ 1 - \left ( \frac{M_{\stau}}{E_{beam}} \right )^2 }
              \right ]
\label{eq:two}
\end{equation}
In addition, independent of the  model-point, the momentum of \textit{ any} jet should be less than  70~\%~$E_{beam}$.
All these conditions benefit from the well-known initial state, the hermeticity of the detectors, and
the clean final state with no pile-up,
at hand at linear $e^+ e^-$ colliders.
Excluding for the conditions for $\tau$-identification, a signal efficiency above 95 \%
is retained after these cuts.

\begin{figure}[t]
  \centering
       \includegraphics [scale=0.25]{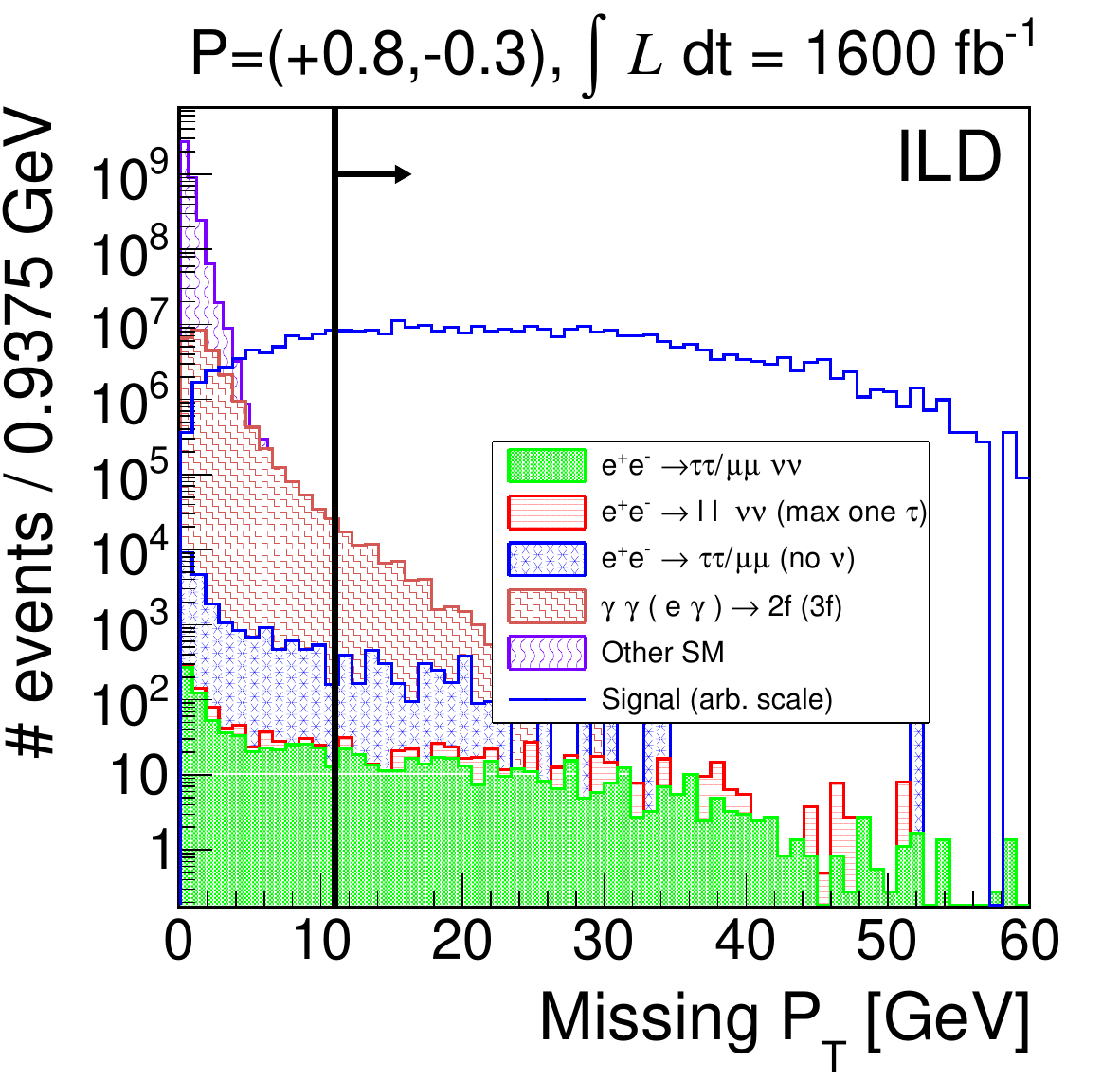}
       \includegraphics [scale=0.25]{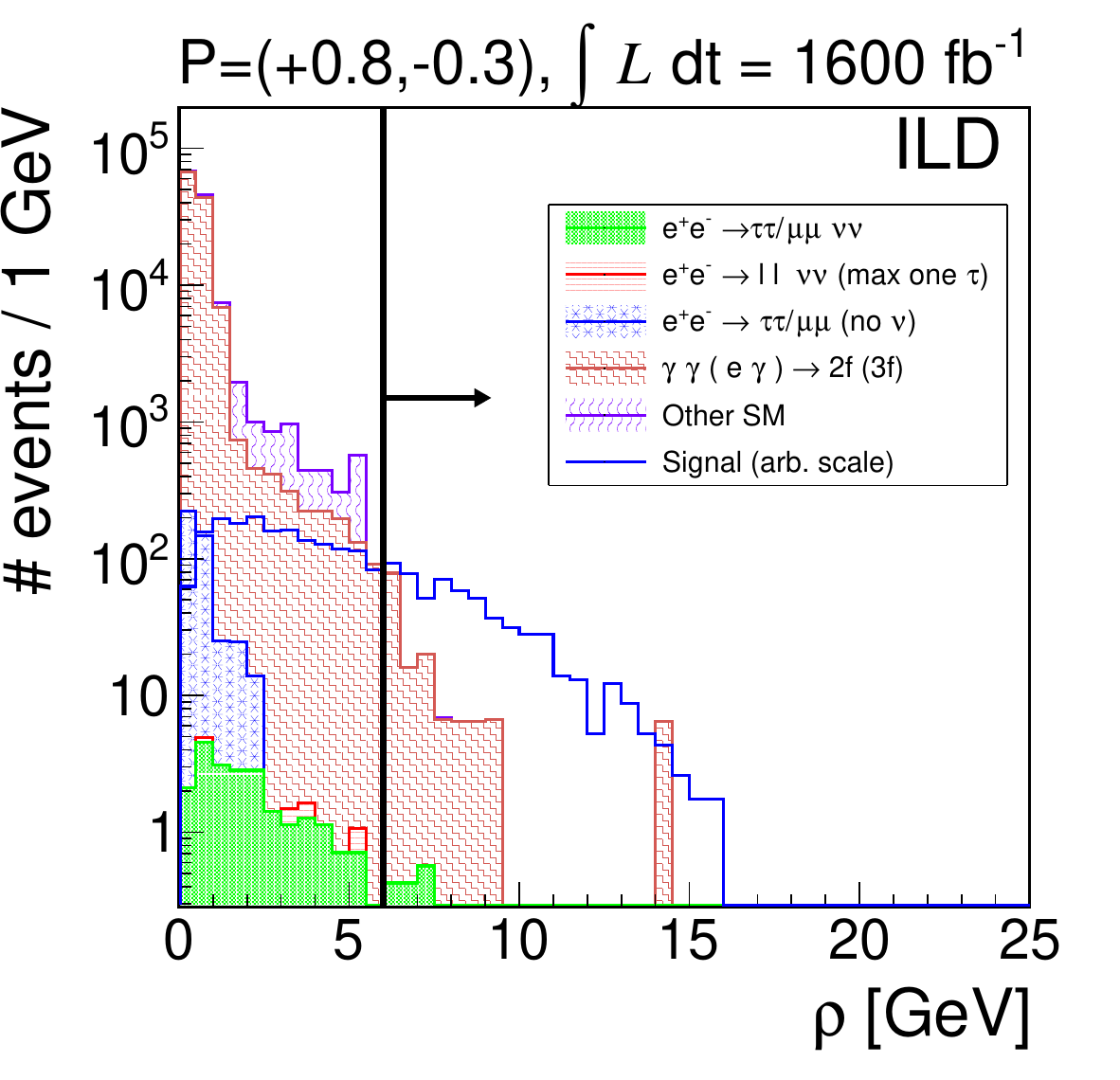}
    \caption{   
     Left:  Distribution of the missing transverse momentum for   $M_{\widetilde{\tau}}$ = 230 GeV $\Delta M$ = 34 GeV,
     Right: Distribution of the variable $\rho$, described in the text, $M_{\widetilde{\tau}}$ = 245 GeV $\Delta M$ = 10 GeV. 
The signals are on arbitrary scale, and
  the arrows indicate the region where events are accepted.\label{fig:cutexamples}}
  \end{figure}

Further selections are based on properties that $\stau$'s might have, but background rarely has.
This includes
high  missing transverse momentum $P_T^{miss}$,
see Figure \ref{fig:cutexamples} (left),
large acoplanarity and high angles to the beam-axis.
An important cut is the one on $\rho$, the $P_T$ with respect to thrust-axis projected on the plane perpendicular to the beam-axis.
$\rho$ will be low
in a $\eeto \tau\tau$ event, or generally any $\tau\tau$ event with $\tau$'s produced back-to-back in
the transversal view, even if the event shows both large acoplanarity and large  $P_T^{miss}$.
This is  because this configuration will only happen if one of the $\tau$'s decays such that most momentum is taken by
  the visible system, while the other does the opposite: most momentum is taken by the neutrino(s).
  This yields large  $P_T^{miss}$ and high acolinearity, but low $\rho$. There is no such correlation for
  $\tau$'s from $\stau$ decays, since the $\tau$'s are \textit{ not} back-to-back, see Figure \ref{fig:cutexamples} (right).

A set of cuts are applied specifically aimed at properties of the irreducible sources of background.
The WW background is highly charge-asymmetric, so a cut on $q_{jet} \cos{\theta_{jet}}$ strongly
reduces this source of background.
The ZZ background tends to have a visible mass in the vicinity of the Z mass, so
this region is also cut out.

Finally, the background still present at this stage often has substantial  energy at
small angles, or contains important energy deposits in isolated neutral clusters,
so these properties are also vetoed.
\subsection*{\textit{ Beam-induced backgrounds}}
  At a linear collider, the $e^+e^-$ beams are accompanied by real  and virtual photons.
The interactions between these produce
  low $P_T$ hadron events.
  At the ILC operating at $E_{CM}$=500 GeV (ILC-500), 1.05 such events are expected on average
  per beam-crossing,
  at CLIC-380(3000), 0.17(3.1) are expected, 
but at  FCCee or CepC, hardly any are.
However,   low $P_T$ hadrons are ``physics'': the total number 
       collected
       scale with the integrated luminosity.
  The photon-photon interactions also create  $e^+e^-$pairs.
  At the ILC, 10$^5$ such pairs are produced per bunch crossing (BX), but  only around ten will hit any tracking detector;
  the vast majority only hits the very forward calorimeter, or escapes down the beam-pipe.
  This background source can be expected to be absent at FCCee.
The $\gamma\gamma$ interactions are independent of the $e^+e^-$ process, but can happen simultaneously to it
(overlay-on-physics events ) or not (overlay-only events).
The overlay-on-physics events will  not be an issue at FCCee, due to low per-BX luminosity.
At the ILC, there is a large effect for low $\Delta M$, but hardly any for
$\Delta M > 10$ GeV, see Figure 
\ref{fig:significancesilcclicunpol} (left)).
    On the other hand, the number of overlay-only events scales with the integrated luminosity, so here
    the effect is the same for the ILC and for the FCCee.
However, the details enter: Smaller beam-spot, trigger-less operation, thinner beam-pipe and vertex detector,
     polarisation, all yields more tools  to the linear options to mitigate this issue.
One will need reduction-factor $\sim 10^{-10}$, which can be shown to be achievable.
  Some slight effect remains at  $\Delta M =2$ - it become completely negligible  with respect to other backgrounds at $\Delta M =10$.



\section{Impact of specific ILC/ILD features}

\subsection*{\textit{ Energy, trigger-less operation}}
It is obvious that energy is a strong advantage for any linear option, compared to circular machines.
An increase in centre-of-mass energy covers much more parameter
          space, up to close to the kinematic limit.
Trigger-less operation of the detectors
is a big advantage when searching for unexpected signatures. 
Such operation is easily feasible at linear colliders due to the low collision frequency,
but not possible at circular colliders.
\subsection*{\textit{ Polarisation}}
\begin{figure}[b]
          \centering
             \includegraphics [scale=0.28]{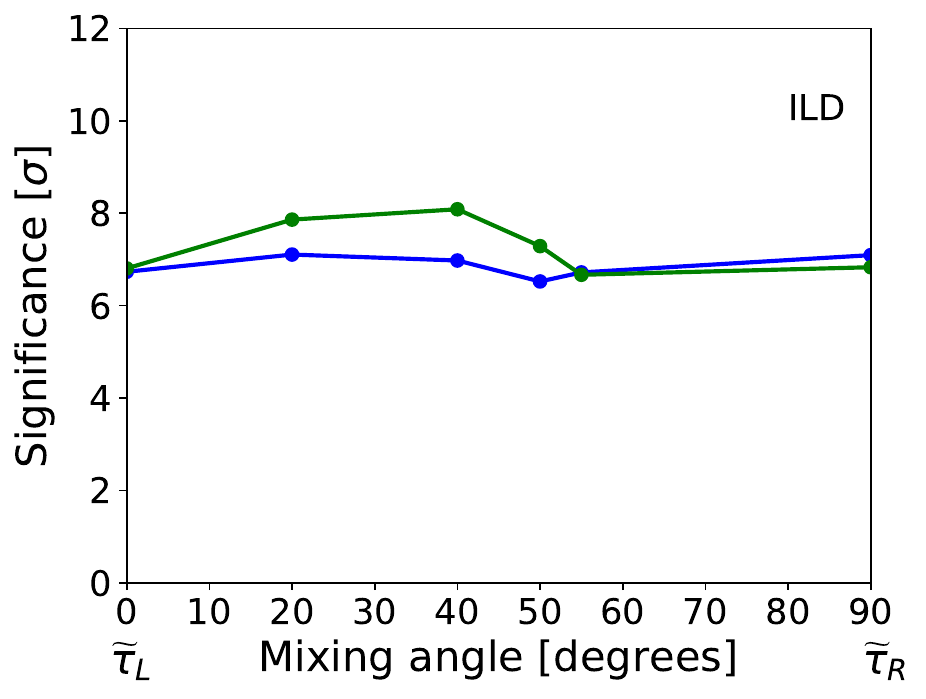}
              \includegraphics [scale=0.28]{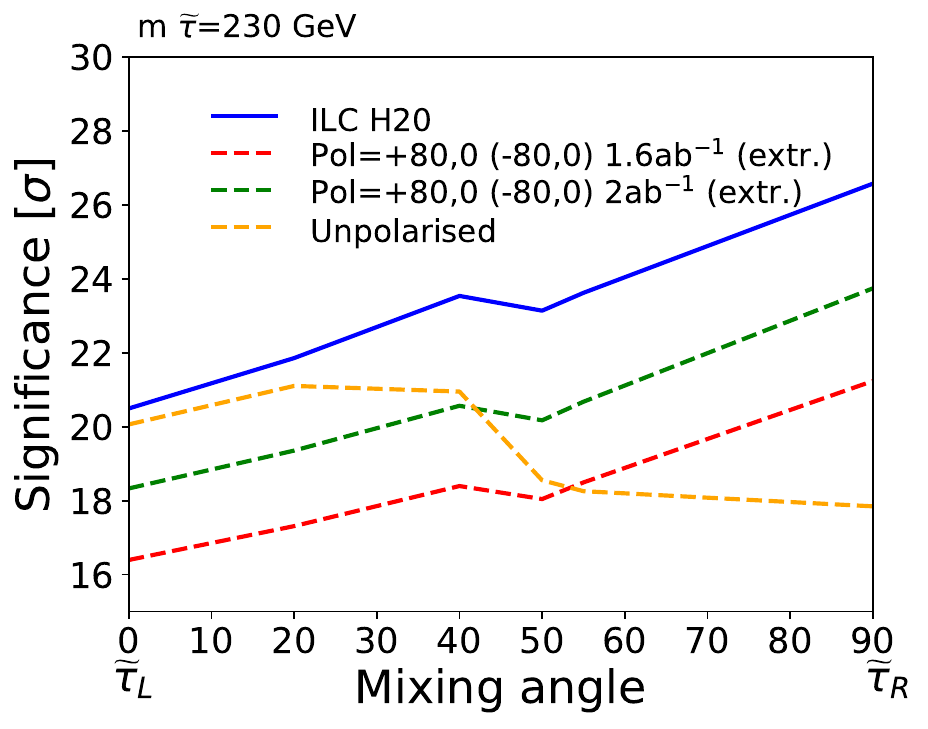}
       \includegraphics [scale=0.20]{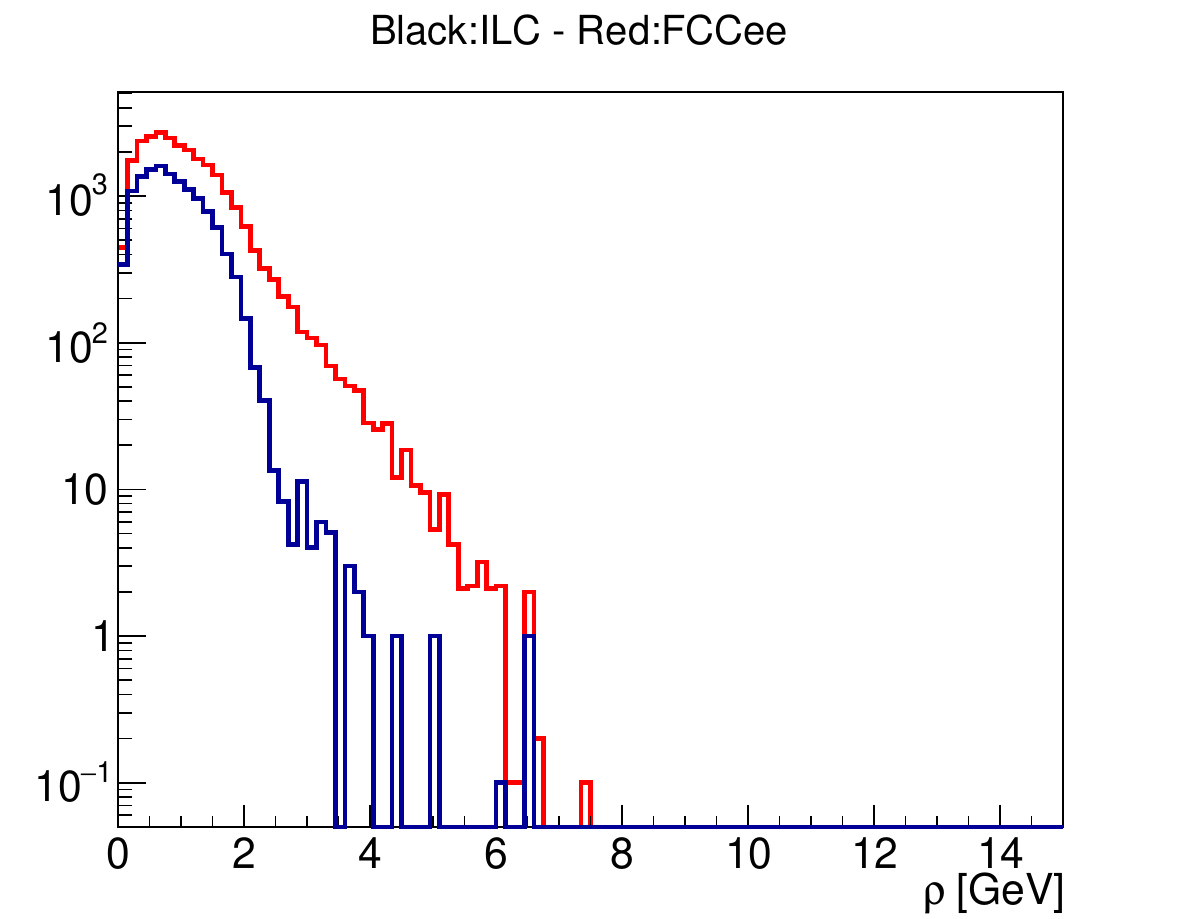}
         \caption{Left: Significance of a $\widetilde{\tau}$ with $M_{\stau}$ = 240 GeV, $\Delta M$ = 10 GeV.
Blue lines correspond to the case with all backgrounds, including overlay tracks, while the green curve corresponds to the study where the overlay tracks are not included.
Middle:  Significance of a $\widetilde{\tau}$ with $M_{\stau}$ = 240 GeV, $\Delta M$ = 10 GeV for different assumptions
on beam polarisation, as indicated by the legend.
Right: The distribution of the $\rho$ variable in $\gamma\gamma$ events after vetoing
signals in the forward calorimeters for ILC conditions (black) and FCCee ones (red). \label{fig:significancesilcclicunpol}}
                  \end{figure}

Beam polarisation allows for
the  combination of different
polarisation samples in such a way that equal sensitivity to any mixing angle
can be achieved.
Control over the beam polarisation also provides overall higher sensitivity since 
            likelihood ratio weighting becomes possible.
If, in addition,  \textit{ both} beams are polarised, the effective luminosity for s-channel 
processes is increased, e.g. an increase by 24 \% for the ILC with respect to a machine none or only one
polarised beam.
 This represents a clear edge for the ILC:  CLIC/C$^3$  only foresees $e^-$ polarisation, FCCee
 none at all. CepC studies if polarisation \textit{ might} be
  possible. See Figure \ref{fig:significancesilcclicunpol} (middle).
\subsection*{\textit{ Luminously, Beam-induced backgrounds }}
High total integrated luminosity is the strong points for FCCee and CepC. 
  However, higher luminosity gives  very little improvement.
   For instance, a change of the total collected data from  2 to 5 (10) ab$^{-1}$ at 250 GeV for $\Delta M$ = 2 GeV
            changes the exclusion limit on $M_{\stau}$ from 112 to 117 (117) GeV, 
            the improvement is negligible for $\Delta M$ = 10 GeV.

For the beam-induced backgrounds, the 
overlay-on-physics background will not be an issue for the circular colliders, due to the low per-BX-luminosity.
The overlay-only background, will, to first order, be similar for both options, since it scales with
the total luminosity.
\subsection*{\textit{ Hermeticity}}
For the hermeticity of the detectors, the issue is if one can detect the beam-remnant $e^{+/-}$ in $\gamma\gamma$ processes.
If not, false missing $P_T$ will be seen.
Here again, the conditions at linear colliders are much more benign than at their circular counter-parts.
The ILD at the ILC is hermetic down to 6 mrad from the beam-axis, 
while any detector at FCCee can only cover the region down to hermetic to 50 mrad to the beam,
due to the real-estate required for the final focus of the beam-delivery system.
For $\stau$, some of the missing $P_T$ is due to the unseen neutrinos
    from the $\tau$-decay, so the effect on the missing  $P_T$ itself is not so drastic.
  However, the $\rho$ variable is designed to see the difference between $\tau$'s 
  that are back-to-back, or not, and becomes much less effective if the hermeticity is compromised.
\subsection{ILC-500 to FCCee-240 comparison}\label{sect:extrap}
It is beyond the scope of this work to make a full study for FCCee-240.
However, some well-founded conclusions can be drawn by extrapolating the
ILC-500 results to FCCee-240 conditions.
This includes both re-scaling the results to a lower $E_{CM}$,
taking the different beam-conditions into account,
and evaluating the effect of the change in detector acceptance.

    For the background, the total measured energy scales up or down
    linearly with $E_{CM}$.
    Away from resonances, the angular distributions do not change with
    $E_{CM}$, so that transverse quantities - or projected ones in any
    direction -
    scales linearly with $E_{CM}$.

  For the signal, the highest possible $P_T$ of any visible decay 
products of the $\stau$ is $ P_{max}$ (eq.~\ref{eq:two}).
So, if one scales both $M_{\stau}$  and $M_{LSP}$ by $E_{beam}$, both brackets remain
unchanged, so that $ P_{T~max}$ scales with $E_{beam}$, just like the background.
    The conclusion is that one expects S/B at one  $E_{CM}$ to be the same as that at another
    $E_{CM}$ if one scales the kinematic cuts and the SUSY masses with the
    ratio of the two
    $E_{CM}$.  
At some distance above threshold, both background and signal would be expected to scale as 1/$E^2_{CM}$,
so both  S and B are 4.3 times higher at 240 GeV compared to 500 GeV.
\footnote{At linear  colliders, the luminosity increases with $E_{CM}$, so one would expect that
the significance (S/$\sqrt{\mathrm{B}}$) of a signal with certain SUSY-masses at one energy, would correspond to
a significance for signal with SUSY-masses scaled by the ratio of the energies would change by a factor 
equal to the square root of this factor, provided the same amount of time are spent at the two energies.}
If S/B is the same, S/$\sqrt{\mathrm{B}}$ is 2.08 times better at 240, but only if the efficiency is the same.

Some of the effects of the various differences between the ILC and the FCCee conditions
can readily be found, by (hypothetically) changing the conditions for the ILC-500
analysis. By removing polarisation from the analysis,
the increase of
effective luminosity is lost, and the possibility to do Likelihood ratio weighting
no longer exists.
As an example, we find that the signal point with $M_{\stau}$ = 245 GeV and
$\Delta M$ = 8 GeV would have an significance of 2.54 $\sigma$ under the ILC conditions,
but only  of 1.8 $\sigma$ for unpolarised beams.
While the absence of overlay-on-physics under FCCee conditions is an advantage
at the very lowest mass-differences, this advantage is no longer present for  
$\Delta M$ = 8 GeV or larger.

To evaluate the effect of lower hermeticity of the detectors at FCCee,
we note that the background at $\Delta M \sim$ 10 GeV is dominated by
the $\gamma\gamma$ background: For $M_{\stau}$ = 245 GeV and
$\Delta M$ = 10 GeV, 215 such background events are expected in the ILD at the ILC
in the case of
unpolarised beams, against only 19 from all other sources.
(The significance of the signal at this point is just above 2 $\sigma$
with unpolarised beams).
We can therefore make an estimate of the increase in background from
modifying the acceptance of the forward calorimetry at generator level,
and for the $\gamma\gamma$ background only.
We find that one would need to increase the cut in $\rho$ by 75 \% to
keep the same level of background from this source.
However, with this modified cut 82 \% of the signal would be lost,
and the significance would go down to only 0.4 $\sigma$,
and S/B would be 2.6 \%.
According to the scaling with  $E_{CM}$ above, the same S/B would be expected for 
the signal point $M_{\stau}$ = 118 GeV and
$\Delta M$ = 4.8 GeV at FCCee-240. Both signal and background cross-sections would be
4.3 times higher, and the significance would be 2.08 times better, i.e. 0.8 $\sigma$.
To reach 2 $\sigma$, even at this more than two times lower $\stau$ mass,
the FCCee would need to collect 6 times more luminosity
than what is foreseen for the ILC at 500 GeV, i.e. 24 ab$^{-1}$, much more than projected for
FCCee, even with four experiments.

\section{Results}

\begin{figure}[t]
  \centering
       \includegraphics [scale=0.38]{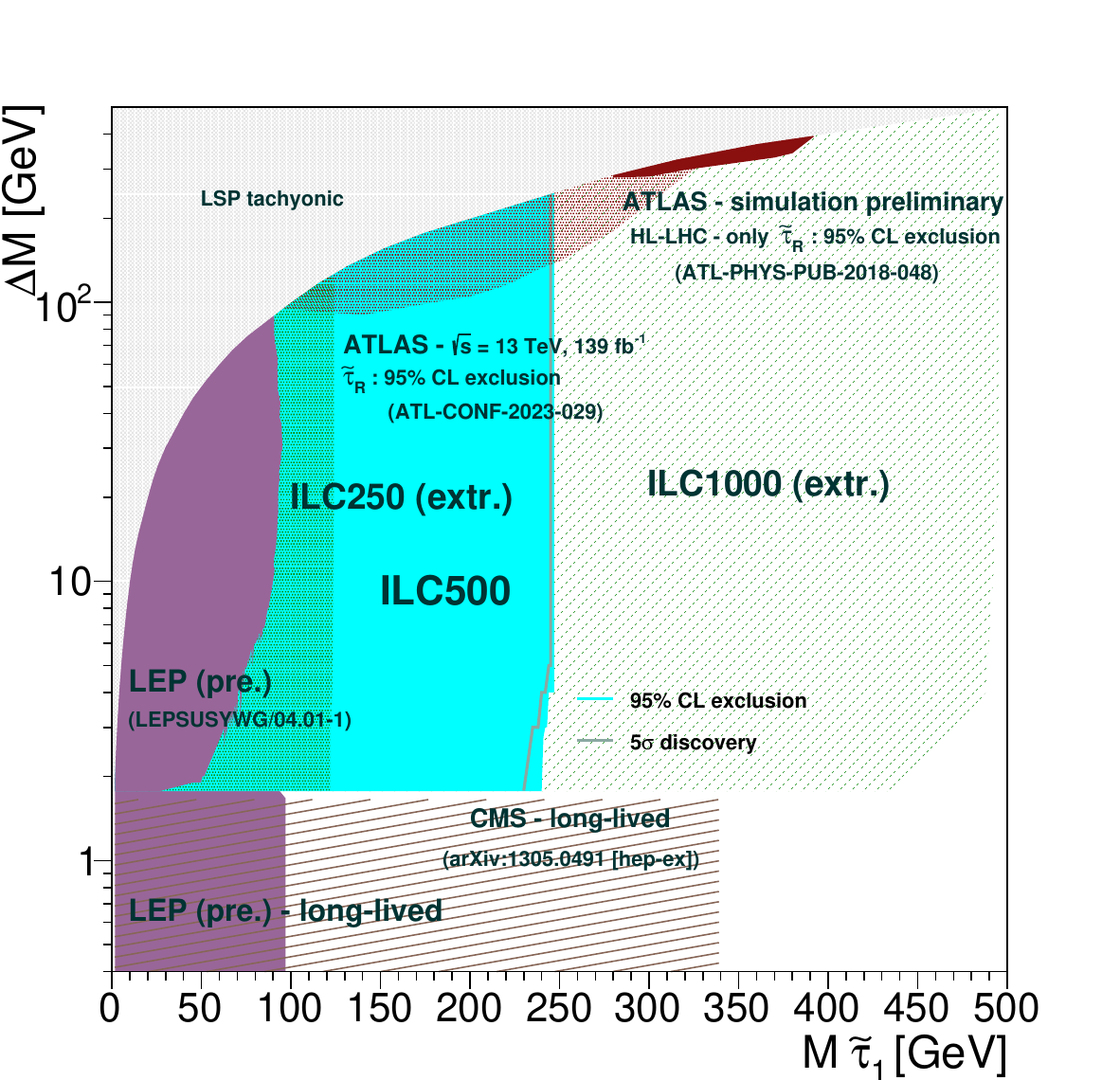}
  
       \caption{In cyan, the exclusion reach  in the $\Delta M$ vs. $M_{\stau}$ plane for $\stau$'s for ILC-500 obtained in this
         study is shown. The discovery reach is shown by the line slightly to the left of the edge of the exclusion region. Also shown is the extrapolations of the current
         study to ILC-250 and ILC-1000, as well as the current
         limits, cf. Figure \ref{fig:currentlimits}.\label{fig:exclusionlimitsfinal}}
  
\end{figure}

The final exclusion limits obtained in this study
are shown
in Figure \ref{fig:exclusionlimitsfinal}, together with the current
limits from LEP and LHC.
Note that at the ILC discovery and exclusion are almost the same,
while the LHC limits and the HL-LHC projections are only exclusion-limits -
no discovery reach is found.
Also shown are the limits the current study would imply for
a ILC-250 and ILC-1000,
using the recipe outlined in section \ref{sect:extrap},
and luminosities as assumes by the H20 scenario.


\section{Conclusions}
  Even after the HL-LHC the $\stau$-LSP mass plane will remain almost completely unexplored.
  Future electron-positron colliders are ideally suited for $\stau$ searches.
  Both the $\stau$ mixing and the nature of the LSP influences the production cross-sections and decay 
            kinematics. In this analysis we made sure that we  studied the ``worst scenario'' taking both of these
into account within a realistic full-simulation of the ILD at ILC-500.
We find that having polarised beams allows for the best exploitation of the data, and that combination of data-taking 
with different signs enables equal 
            sensitivity to all mixing angles.
We also find that beam-induced backgrounds at Linear Colliders can be mitigated up to small residual 
            impact of $\sim$ 1GeV on highest reachable mass for lowest $\Delta M$.
 Higher centre-of-mass energies cover much more parameter space than what higher luminosity would give. 
            For instance, an increase of ILC-250 luminosity from 
            2 to 10 ab$^{-1}$ only affects the $\stau$ mass limit  by 5 GeV.
 Finally, by comparing with the case of an detector at FCCee, we find that the hermeticity of detector is crucial.
This implies that at circular colliders, at most some modest amelioration of the limits from LEP can be expected.

\section{Acknowledgements}
We would like to thank the LCC generator working group and the ILD software
working group for providing the simulation and reconstruction tools and
producing the Monte Carlo samples used in this study.
This work has benefited from computing services provided by the ILC Virtual
Organisation, supported by the national resource providers of the EGI
Federation and the Open Science GRID.

\printbibliography
\end{document}